\documentclass[12pt]{article}
\usepackage{graphicx}
\usepackage{natbib} %comment out if you do not have the package
\usepackage[a4paper]{geometry}
\usepackage{graphicx}
\usepackage[textwidth=8em,textsize=small]{todonotes}
\usepackage{amsmath}
\usepackage{amssymb}
\usepackage{bbm}
\usepackage{array}
\usepackage{booktabs}
\usepackage{physics}
\usepackage{diagbox}
\usepackage{url}
\usepackage{appendix}
\graphicspath{{Figures/}}
\usepackage{url} % not crucial - just used below for the URL 

\newcommand{\blind}{0}

\begin{document}

\def\spacingset#1{\renewcommand{\baselinestretch}%
{#1}\small\normalsize} \spacingset{1}

%%%%%%%%%%%%%%%%%%%%%%%%%%%%%%%%%%%%%%%%%%%%%%%%%%%%%%%%%%%%%%%%%%%%%%%%%%%%%%

\if0\blind
{
  \title{\bf A Multi-Site Stochastic Weather Generator for High-Frequency Precipitation Using Censored Skew-Symmetric Distribution}
  \author{Yuxiao Li and Ying Sun\thanks{
   This research was supported by King Abdullah University of Science and Technology (KAUST), Office of Sponsored Research (OSR) under Award No: OSR-2019-CRG7-3800.}\hspace{.2cm}\\
    Statistics Program, King Abdullah University of Science and Technology}
    %and \\
    %Ying Sun \\
    %Statistics Program, King Abdullah University of Science and Technology}
  \maketitle
} \fi

\if1\blind
{
  \bigskip
  \bigskip
  \bigskip
  \begin{center}
    {\LARGE\bf A Multi-Site Stochastic Weather Generator for High-Frequency Precipitation Using Censored Skew-Symmetric Distribution}
\end{center}
  \medskip
} \fi

\bigskip
\begin{abstract}
Stochastic weather generators (SWGs) are digital twins of complex weather processes and widely used in agriculture and urban design. Due to improved measuring instruments, an accurate SWG for high-frequency precipitation is now possible. However, high-frequency precipitation data are more zero-inflated, skewed, and heavy-tailed than common (hourly or daily) precipitation data. Therefore, classical methods that either model precipitation occurrence independently of their intensity or assume that the precipitation follows a censored meta-Gaussian process may not be appropriate. In this work, we propose a novel multi-site precipitation generator that drives both occurrence and intensity by a censored non-Gaussian vector autoregression model with skew-symmetric dynamics. The proposed SWG is advantageous in modeling skewed and heavy-tailed data with direct physical and statistical interpretations. We apply the proposed model to 30-second precipitation based on the data obtained from a dense gauge network in Lausanne, Switzerland. In addition to reproducing the high-frequency precipitation, the model can provide accurate predictions as the long short-term memory (LSTM) network but with uncertainties and more interpretable results.
\end{abstract}

\noindent%
\vfill

\newpage
\spacingset{2} % DON'T change the spacing!
\section{Introduction}\label{sec:intro}

Tremendous efforts have been made to model, forecast, and reproduce local and global precipitation patterns. Among these efforts, the stochastic weather generator (SWG) makes use of statistical tools to simulate random sequences and reproduce atmospherical variables efficiently \citep{wilks1999weather}. Typically, an ideal stochastic precipitation generator (SPG) should be able to reproduce the statistical properties of occurrence, intensity, and dry spell length of precipitation. The idea of reproducing physical processes with a virtual replica applied in other areas, such as ``digital twin'' \citep{boschert2016digital, soderberg2017toward, el2018digital}. A Digital twin integrates many modern techniques such as internet of things and artificial intelligence. It is also the critical sector of smart city  \citep{falconer2012smart} and industry 4.0 \citep{gilchrist2016industry}.

SPGs have proven to be important in water resources research. As an example of digital twins, SPGs provide valuable information for water resource management for a smart city \citep{parra2015development}. In hydrologic and agricultural science, SPGs can serve as an input in further simulations of erosion, flood and crop growth \citep{mary2009conceptual}.  Moreover, as the primary atmospheric variable in SWGs, precipitation is typically used to generate other variables, such as wind and solar irradiance, due to their close association with rainfall occurrence \citep{richardson1981stochastic}. Techniques in modeling precipitation data can be also beneficial to other fields of study, such as sociology \citep{heckman1976common} and economics \citep{mcdonald1980uses}, where the data properties are often similar to those of precipitation, e.g., they are zero-inflated, nonnegative, right-skewed, heavy-tailed, and correlated in space and time.

Due to their wide applicability and the intriguing challenges, studies of SPGs have drawn attentions since 1960s \citep{gabriel1962markov} and been systematic reviewed in \citet{wilks1999weather}, \citet{srikanthan2001stochastic}, and \citet{ailliot2015stochastic}. Traditional studies mainly focus on the SPGs with a low temporal resolution, usually on a daily scale, due to the data available. For instance, the most popular chain-dependent model \citep{katz1977precipitation, richardson1981stochastic} assumes that the occurrence processes can be modeled independently of intensity by Markov chains, and the intensity processes can be estimated conditionally on wet events using Gamma or exponential distributions. However, their assumption of independent occurrence is not appropriate for high-frequency precipitation, since greater quantities of rainfall in the past may lead to a significantly higher probability of occurrence \citep{koch2015frailty}. 

\begin{figure}[h]
\centering
~~~~~~~~
         \includegraphics[width=.8\linewidth]{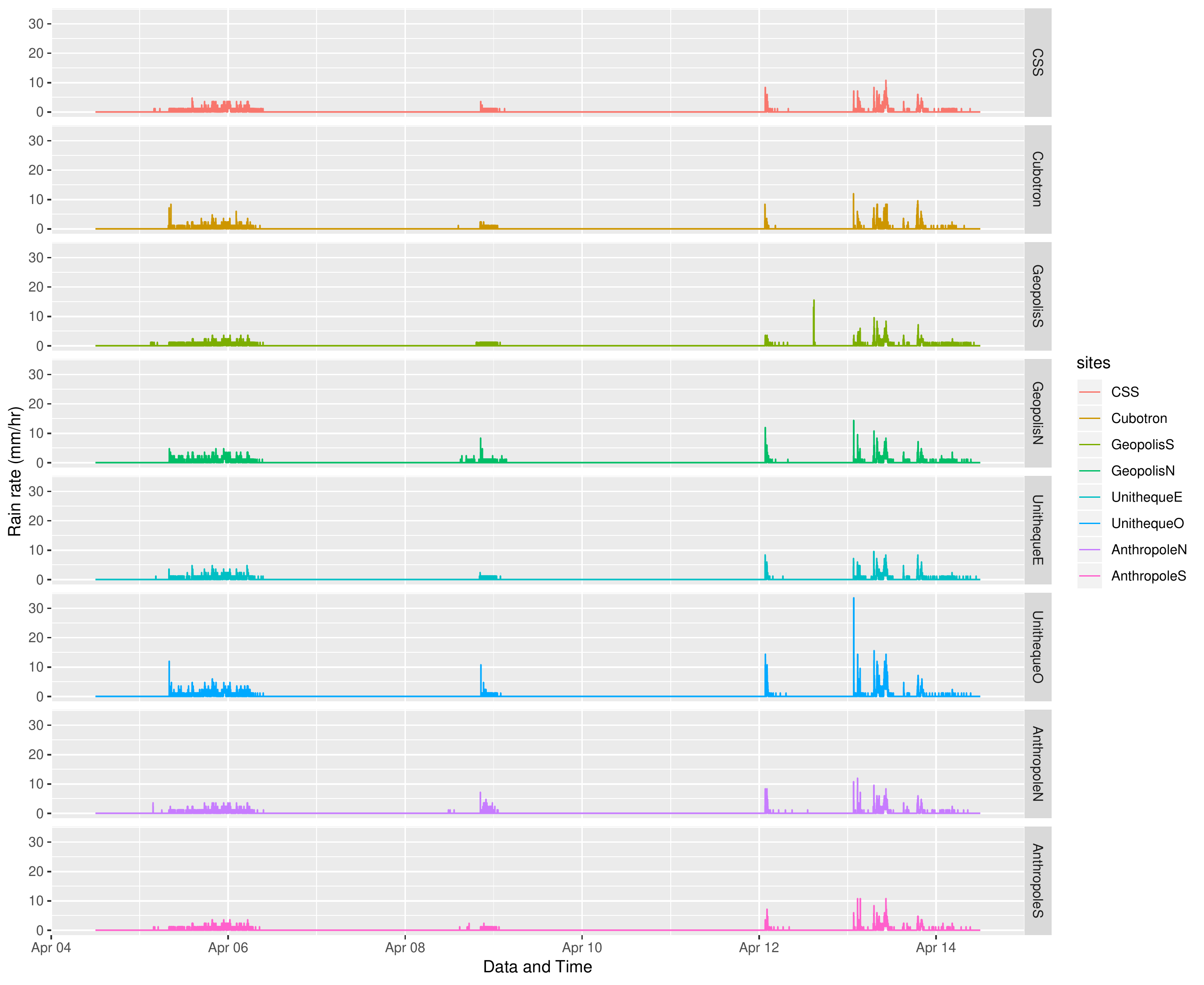}
         \caption{The eight time series of rain rate (unit: mm/hr) collected by Dryptich Pluvimate acoustic rain gauges every 30 seconds from April 4th to April 14th, 2016.}
         \label{fig:rain_raw}
\end{figure}

Recently, acoustic rain gauges are able to provide precise and high-frequency data which are undetectable by the classical measurement devices, such as tipping-bucket rain gauge, satellite-based radar, and terrestrial radar. The high-frequency dataset is valuable to many rainfall-related phenomena, such as rapid surface water runoff, flash flooding, and small river catchments \citep{chan2016characteristics}.
However, the rainfall model at this scale has been rarely developed and assessed. To our knowledge, only \citet{benoit2018stochastic} has performed an investigation of the minutely and sub-kilometer SPGs with the same device; in their study, the spatial dependence was the major concern.  Even though broad literatures exist on the development of hourly and daily SPGs, we cannot assume that their results can extend to our timescales. In this study, our objective is to use this single model to reproduce precipitation on a very fine scale both spatially (within one radar pixel, $10-100$m) and temporally (less than a minute). The data of our interest contain 30-second rainfall that were collected on eight acoustic rain gauges located within a radius of 1 km in the University of Lausanne campus in Switzerland (see Figure \ref{fig:rain_raw}).

To handle the high-frequency data, the common practice is to make use of censored models to drive both the occurrence and the intensity processes. The models are also called truncated models by \citet{allard2012modeling}, and the Tobit models \citep{mcdonald1980uses} from econometrics.  In the censored models, both the occurrence and intensity are driven by uncensored processes, where the dry events are zero values left-censored at a certain threshold, and the wet events are modeled by the process with a positive support. 

Meta-Gaussian processes are the most popular way for the spatio-temporal dependence and non-Gaussian features of rainfall data \citep{ailliot2009space, kleiber2012daily, baxevani2015spatiotemporal}. Although the methodologies provide high flexibility with regard to the spatio-temporal dependence, several limitations remain. First, the transformation of Gaussian processes is ad-hoc, ranging from simple power or logarithm transformations  \citep{bell1987space,glasbey1997rainfall,durban2001weather} to complex Tukey g-and-h or hybrid Gamma transformations \citep{baxevani2015spatiotemporal, xu2017tukey}. The efforts to achieve normality can be tedious. Second, the latent Gaussian processes make the interpretation difficult, especially when different transformations are suggested \citep{ailliot2015stochastic}. Third, Gaussian processes are more suitable for denser spatial data rather than spatially sparse data, such as the multi-site high frequency time series.

Another popular way to quantify the spatio-temporal dependence in SPG is the vector autoregressive (VAR) processes \citep{hamilton1994time,sigrist2012dynamic, rasmussen2013multisite, koch2015frailty}. The VAR framework has been widely used in a set of high-frequency data such as stock returns and brain signals. For precipitation, the state-of-arts censored VAR model is proposed by \citet{koch2015frailty}, for which the interpretation is easy and its likelihood-based inference is straightforward. However, this model only considers heteroscedastic Gaussian and meta-Gaussian error, which may not be adequate for data with skewed or heavy tailed distributions.
% and the VAR framework can apply to the data process directly. 
%Although the VAR framework is less flexible in spatial domain, its likelihood-based inference is straightforward. 
%In addition, this VAR model is similar to the frailty-contagion model from finance \citep{azizpour2008self}, which is suitable for high-frequency data analysis.

Therefore, we propose to use the skew-symmetric families \citep{azzalini1985class} in the censored VAR model. The skew-symmetric distributions have been generalized and applied in many fields of study \citep{azzalini1999statistical, genton2004skew, azzalini2013skew} in modeling the obviously non-Gaussian features, but rarely used in the SWG. The only application to SWG was investigated by \citet{flecher2010stochastic}, where the multivariate skew-normal distribution was adopted to model multiple atmospheric variables. 

As an example, the density function of a univariate skew-t random variable is specified as 
\begin{equation}
f_{\mathcal{ST}}(x;\xi,\omega,\alpha,\nu)=2t(u;\nu)T(\alpha u\sqrt{(\nu+1)(\nu+u^2)};\nu+1), u=\frac{x-\xi}{\omega}, 
\label{eq:st}
\end{equation}
where $\xi$ and $\omega$ are the location and scale parameters, $\alpha$ is the skewness parameter, $\nu$, is the degree of freedom, $t(\cdot)$ is the standard t density function, and $T(\cdot)$ is the standard t distribution function.

The skew-normal distribution is a special case of the skew-t distribution when $\nu=\infty$. Similar to the normal or student-t distributions, $x$ can take any real value. In contrast, two extra parameters in the skew-t distribution controls the skewness and tail behavior. Another attractive feature of the skew-t distribution is the stochastic representation \citep{azzalini2012some}. A skew-t random variable $X$ can be expressed by hidden selective mechanism such that $X=(X_1|X_2>0)$, where $X_1$, $X_2$ are correlated t random variables with the same degree of freedom.

Since the generation mechanism behind the precipitation can be viewed as a hidden selection process, using a skew-symmetric distribution in modeling precipitation data produces nice physical interpretations, as will be further explained in Section~\ref{sec:inter}. By incorporating skew-symmetric distributions and a censored VAR framework, we propose a new stochastic precipitation generator that 

1) uses a single spatio-temporal model to simultaneously drive the precipitation occurrence and its intensity;

2) has direct interpretation to the precipitation process;

3) allows for flexible and tractable tail behaviors, which is crucial in modeling high-frequency precipitation data;

4) implies parsimonious parametrization and efficient data generation.

%Our objective is to reproduce precipitation at a very fine scale that spatially within one radar pixel (10-100m) and temporally less than one minute, motivated by the 30-sec rainfall data collected by 8 Pluvimates acoustic rain gauges set up on the campus of the University of Lausanne.  These recently designed acoustic rain gauges with the detailed description in  provide higher precision and resolution than usual tipping bucket gauges. The fine resolution allows us to capture the small rain cells, undetectable on other measurements such as satellite-based radars, terrestrial radars, and tipping bucket. 

%Here, first add a brief introduction of the multivariate skew-symmetric distribution (or univariate if it is easier to explain), and hidden selection. You can already use pdf or notations, for example, which parameter controls the skewness and which controls the tail, etc.

%Then say why it is appealing in precipitation modeling, followed by the comments about the hidden selection representation which allows for nice physical interpretations.
For the application to high-frequency precipitation data, recent studies also propose to use deep learning methods such as long short-term memory (LSTM) or other recurrent neural networks (RNN) \citep{xingjian2015convolutional, shi2017deep}. Although SPG focuses on reproducing the precipitation in the functional perspective, it is also important to show the results of prediction, especially comparing to these deep learning methods. We choose the multivariate LSTM networks as our competing methods to show the difference between our stochastic generators and the deep learning forecasters.

The rest of our paper is organized as follows. In Section \ref{sec:model}, we introduce the new class of SPGs, provide model properties, and describe the inference procedure. In Section \ref{sec:sim}, we present simulation studies to validate our inference method and compare its performance to other models. In Section \ref{sec:app}, we show the performance of the proposed SPGs on the high-frequency rainfall dataset collected at the University of Lausanne campus. In Section \ref{sec:dis}, we summarize our main results and discuss the potential limitations.

%A network consists of 8 Driptych Pluvimates set up on the campus of the University of Lausanne. The Driptych Pluvimates acoustic rain gauges is able to count individual drops with 30 seconds resolution. Such precision and temporal resolution allow capturing the passage of small rain cells, invisible on the radar. 

%%%%%%%%%%%%%%%%%%
%                  Model
%%%%%%%%%%%%%%%%%%

\section{Censored skew-symmetric VAR models}\label{sec:model}
\subsection{Censored VAR models}
Let $Y_t(s)$ be the precipitation amount observed at a site $s$ and time $t$,  $s=1,\dots,N, t=1,\dots,T$. Collect $Y_t(s)$ as an $N\times1$ vector $\textbf{Y}_t$. Then we specify the multi-site precipitation generator based on censored VAR as 
\begin{equation}
{Y}_t(s)=\begin{cases}
\begin{array}{c}
\boldsymbol{\beta}_{s}'\textbf{Y}_{t-1}+{\varepsilon_t}(s),\\
0,
\end{array} & \begin{array}{c}
\boldsymbol{\beta}_{s}'\textbf{Y}_{t-1}+{\varepsilon_t}(s)>{u}_t(s),\\
\boldsymbol{\beta}_{s}'\textbf{Y}_{t-1}+{\varepsilon_t}(s)\leq {u}_t(s),
\end{array}\end{cases}
\label{eq:pg}
\end{equation}
where $\boldsymbol{\beta}_{s}$ is an $N\times1$ vector of autoregression coefficients for site $s$, ${u}_t(s)$ is the space-time varying cutoff vector representing the censoring threshold associated with the rain probability, and ${\varepsilon_t}(s)$ is the random error.

The censored VAR generator in \eqref{eq:pg} is originally proposed by \citet{koch2015frailty}. The model \eqref{eq:pg} differs from other SPGs since it avoids applying any ad-hoc transformation to the rainfall intensity when ${Y}_t(s)>0$. Therefore, the likelihood of ${Y}_t(s)$ has an explicit form and parameters in the model have direct interpretation to ${Y}_t(s)$.

The model \eqref{eq:pg} also differs from classical VAR model \citep{sims1980macroeconomics, hamilton1994time}, where ${Y}_t(s)$ is not censored and the error term $\varepsilon_t(s)$ is assumed to be white noise with zero mean and constant variance. Here, a more general case is considered for $\varepsilon_t(s)$ by letting $\varepsilon_t(s)=\sigma_t z_t(s)$. In this setting, $z_t(s)$ are independent, but the error terms $\varepsilon_t(s)$ are not independent in general. 
%%%%%%%%%%%%%%%%%%
%                  Subsection Our Model
%%%%%%%%%%%%%%%%%%

\subsection{Censored VAR models with skew-symmetric errors}
The independent random variables $z_t(s)$ in \citet{koch2015frailty} is assumed to follow a standard normal distribution and the standard deviations are modeled with other atmospheric explanatory variables by a linear regression. However, these explanatory variables are often either unavailable or hard to choose in practice.  This issue becomes more problematic when they assume that $z_t(s)$ is normally distributed because the right-skewed and heavy-tailed features of the rainfall data can be only explained by $\sigma_t$. Then the entire dynamics of the SPG will be heavily influenced by the selected explanatory variables. 

Therefore, we consider another random error that is flexible with the skewness and tail behavior so that the explanatory variables are not required.
Instead of a normal distribution, we assume that $z_t(s)$ follows a family of skew-symmetric distributions \citep{azzalini2013skew} with zero mean and unit variance. As we mentioned in \eqref{eq:st}, $z_t(s)$ itself already describes the skewed and heavy-tailed features, rather than relying on the other explanatory variables.

For the standard deviations, we allow for heteroscedasticity with a temporally varying standard deviation $\sigma_t$. The idea of heteroscedasticity is widely used to predict high-frequency data such as wind power and stock-returns \citep{engle1982autoregressive, bollerslev1986generalized, taylor2009wind}, such as the (generalized) autoregressive conditionally heteroscedastic (ARCH/GARCH) models. Specifically,  we assume that $\sigma_t=b_0+b_1\bar{Y}_{t-1}$, where $\bar{Y}_{t-1}=\sum\limits_{s=1}^N{Y}_{t-1}(s)/N$ and $b_0, b_1\geq0$ to avoid a negative variance. 

To describe the distribution of $z_t(s)$, three candidates in the skew-symmetric family are commonly used: the skew-normal distribution \citep{azzalini1985class}, the skew-t distribution \citep{branco2001general,azzalini2003distributions}, and the skew-Cauchy distribution \citep{behboodian2006new}. Since the skew-t distribution with the degree of freedom $\nu$ includes the skew-normal and the skew-Cauchy distribution as special cases, hereafter we mainly discuss the properties of the skew-t distribution. The results from the skew-normal and skew-Cauchy distributions can be simply obtained by replacing $\nu$ with $\infty$ and $1$, respectively.

We assume that $z_t(s) \overset{\mathrm{iid}}{\sim}\mathcal{ST}(\xi,\omega, \alpha,\nu)$ with density function \eqref{eq:st}. For the skew-t distribution, $\xi$ is not the mean and $\omega$ is not the standard deviation. Instead, $\mathbb{E}\{z_t(s)\}=\xi+\omega b_{\nu}\delta$ and $\mathrm{var}\{z_t(s)\}=\omega^2\left\{\frac{\nu}{\nu-2}-(b_{\nu}\delta)^2\right\}$, where $b_{\nu}=\frac{\sqrt{\nu}\Gamma(\frac{1}{2}(\nu-1))}{\sqrt{\pi}\Gamma(\frac{1}{2}\nu)}$ and $\delta=\frac{\alpha}{\sqrt{1+\alpha^2}}$. To obtain zero mean and unit variance, we set $\omega=\frac{1}{\sqrt{\frac{\nu}{\nu-2}-(b_{\nu}\delta)^2}}$ and $\xi=-\frac{b_{\nu}\delta}{\sqrt{\frac{\nu}{\nu-2}-(b_{\nu}\delta)^2}}$. Therefore, the scaled skew-t distribution only depends on the skewness parameter $\alpha$ and the degree of freedom $\nu$. The corresponding density function and distribution are denoted by $f_{\mathcal{SST}}(\cdot)=f_{\mathcal{SST}}(x;\alpha,\nu)$ and $F_{\mathcal{SST}}(\cdot)=F_{\mathcal{SST}}(x;\alpha,\nu)$, respectively.

%%%%%%%%%%%%%%%%%%
%    Subsection Interpretation
%%%%%%%%%%%%%%%%%%
\subsection{Model implications and interpretations}\label{sec:inter}
The proposed model is flexible and all the parameters in model \eqref{eq:pg} have natural interpretations. For example, a higher $\nu \in \mathbb{R}^+$ and $\alpha \in \mathbb{R}$ imply a lighter tail and larger right-skewness, respectively; $\sigma_t$ introduces the heteroscedasticity; the autoregression matrix $B=(\boldsymbol{\beta}_s)_{s=1}^N=(\beta_{ij})_{N\times N}$ controls the spatio-temporal dependence, and $u_t(s)$ is the threshold that determines the wet or dry probability.

We can derive several important precipitation probabilities conditional on previous observations. First, the conditional dry probability at site $s$ and time $t$ is 
 \[{\footnotesize
\mathbb{P}\{Y_t(s)=0|\textbf{Y}_{t-1}=\textbf{y}_{t-1}\}=
\mathbb{P}\{\varepsilon_t(s)\leq u_t(s)- \boldsymbol{\beta}'_{s}\textbf{y}_{t-1}\}=F_{\mathcal{SST}}\left(\frac{u_t(s)- \boldsymbol{\beta}'_{s}\textbf{y}_{t-1}}{b_0+b_1\bar{Y}_{t-1}}\right).}\]
Hence, a higher dry probability can be reached by either decreasing $b_0,b_1$ or $\alpha$, or by increasing $u_t(s)$ or $\nu$. In particular, if $\textbf{Y}_{t-1}=\textbf{0}$, then we have the consecutive dry probability  $\mathbb{P}(Y_t(s)=0|\textbf{Y}_{t-1}=\textbf{0})=F_{\mathcal{SST}}\left(\frac{u_t(s)}{b_0}\right)$. 

Since the random variables $z_t(s)$ are independent, the simultaneously dry probability at multiple sites is the product of marginal probabilities. Therefore, once we plug in the estimated parameters for those probabilities, which are conditional on previous events, we can immediately obtain the dry/wet probability (rainfall occurrence), consecutive dry/wet probability (distribution of the dry/wet spell length), and the simultaneous dry/wet probability for multiple sites (rainfall spatial pattern).

%We can also derive the associated properties of rainfall intensity such as the $r$-th moment of $Y_t(s)|\textbf{Y}_{t-1}=\textbf{y}_{t-1}$. The results are less trivial, since the $r$-th moment of the truncated skew-t random variable is required \citep{jamalizadeh2009truncated}. The results can be found in the Proof 1 of Appendix.

The SPG in model \eqref{eq:pg} possesses both flexible statistical properties and nice physical interpretations. We illustrate these by representing model \eqref{eq:pg} as a state-space model, i.e., a two-layer model where the transition from zero to positive values is driven by the selection mechanism of the skew-t distribution. The equivalent model can be specified as:

\noindent\begin{minipage}{.45\linewidth}
\begin{equation}
{Y}_t(s)=\begin{cases}
\begin{array}{c}
{X}_{t}(s),\\
0,
\end{array} & \begin{array}{c}
{X}_{t}(s)>{u}_{t}(s),\\
{X}_{t}(s)\leq {u}_{t}(s),
\end{array}\end{cases}
%\tag{3}
\label{eq:stoc1}
\end{equation}
\end{minipage}%
\begin{minipage}{.55\linewidth}
\begin{equation}
{X}_t(s)=g(\textbf{X}_{t-1})+\begin{cases}
\begin{array}{c}
{Z}_t(s),\\
-{Z}_t(s),
\end{array} & \begin{array}{c}
{W}_t(s)>0 ,\\
{W}_t(s)\leq 0,
\end{array}\end{cases}
%\tag{4}
\label{eq:stoc2}
\end{equation}
\end{minipage}
\vspace{5mm}

\noindent where the threshold ${u}_t(s)$ is deterministic, the latent autoregressive process ${X}_t(s)$ depends on a function of $\textbf{X}_{t-1}$ called $g(\textbf{X}_{t-1})$, and the random errors are controlled by two processes, ${Z}_t(s)$ and ${W}_t(s)$. The proof for the equivalence between Equation \eqref{eq:pg} and Equations \eqref{eq:stoc1} and \eqref{eq:stoc2} is given in the Supplements.

In meteorology, it is well known that the precipitation comes from condensed atmospheric water vapor, which is formed at certain temperatures and moisture conditions, and then falls as observable rainfall due to gravity. Our equations  \eqref{eq:stoc1} and  \eqref{eq:stoc2} describe this physical process. 

Equation \eqref{eq:stoc1} is the measurement equation, which we call the ground layer model. It describes the amount of condensed atmospheric water vapor $X_t(s)$ that becomes precipitation $Y_t(s)$ at time $t$ and location $s$. The wet-dry threshold ${u}_t(s)$ represents the necessary conditions for rainfall, defined as the minimum condensed water vapor required for observable rainfall, i.e., to reach the detection limit of the measuring instrument. Equation \eqref{eq:stoc2} is the transition equation, which we call the atmospheric layer model. It describes the formation of condensed water vapor in the atmosphere. The current condensed water vapor $X_t(s)$ is modeled using past observations at all locations $g(\textbf{X}_{t-1})$, with random fluctuations ${Z}_t(s)$ that represent the new formation and dissolution of  condensed water vapor. The fluctuation ${Z}_t(s)$ is not just the symmetric random noise, but is driven by a hidden selection process ${W}_t(s)$, which represents certain meteorological conditions, known as weather fronts such as temperature and moisture. Although the distributions of their elements, $Z_t(s)$ and $W_t(s)$, are both symmetric, when they are correlated, the distribution of $X_t(s)$ becomes skewed. This representation explains that our model is suitable for data that are censored and skewed. Therefore, the proposed SPG can potentially simulate realistic precipitation observations.

%%%%%%%%%%%%%%%%%%
%    Subsection   Inference
%%%%%%%%%%%%%%%%%%

\subsection{Inference and computational issues}\label{sec:inf}
%Although the SPG as a stochastic state-space model shown in Equations \eqref{eq:stoc1} and \eqref{eq:stoc2} has attractive properties, its inference is difficult because the model is non-linear and non-Gaussian. In contrast, the VAR representation in Equation \eqref{eq:pg} belongs to the generalized Tobit model \citep{mcdonald1980uses}. \citet{koch2015frailty} shows that the inference of Equation \eqref{eq:pg} can be achieved by maximizing the likelihood function under Gaussian error. However, in the appendix, we show that the likelihood also applied for other errors. Specifically, 

The inference of model \eqref{eq:pg} is straightforward as it belongs to the generalized Tobit model \citep{mcdonald1980uses}. We show in the supplements that the log-likelihood, $\ell(\boldsymbol{\theta}|\textbf{y}_{1},\dots,\textbf{y}_{T})$, can be written as 
\begin{equation}
{\footnotesize
\begin{array}{r@{}l}
\ell(\boldsymbol{\theta}|&\textbf{y}_{1},\dots,\textbf{y}_{T})=\\&\sum_{t=2}^T\sum_{s=1}^N  \mathbbm{1}_{\{y_t(s)>0\}}\left[\log\left\{f_{\mathcal{SST}}\left(\frac{y_t(s)-\boldsymbol{\beta}_{s}'\textbf{y}_{t-1}}{b_0+b_1\bar{Y}_{t-1}}\right) \right\}-\log\left\{b_0+b_1\bar{Y}_{t-1}\right\}\right] \\
& +\sum_{t=2}^T\sum_{s=1}^N\mathbbm{1}_{\{y_t(s)=0\}} \log\left\{F_{\mathcal{SST}} \left(\frac{{u}_t(s)-\boldsymbol{\beta}_{s}'\textbf{y}_{t-1}}{b_0+b_1\bar{Y}_{t-1}}\right) \right\},
\end{array}
\label{eq:lik}
}
\end{equation}
where $\boldsymbol{\theta}$ is the vector of parameters to be estimated, $\mathbbm{1}_{\{y_t(s)>0\}}$ is an indicator function that takes a value of $1$ when $y_t(s)>0$ and $0$ otherwise, $f_{\mathcal{SST}}(\cdot;\alpha,\nu)$ is the scaled skew-t density function, and $F_{\mathcal{SST}}(\cdot;\alpha,\nu)$ is the distribution function.

To achieve a robust estimation of the unknown parameters with some meaningful interpretations, some of the values in \eqref{eq:lik} are further parameterized.
%First, we estimate the autoregression matrix $B=(\beta_{ij})_{N\times N}$ with space-time varying cutoffs $\{u_{t}(s)\}_{T\times N}$, the standard deviation parameters $b_0$ and $b_1$, the skewness $\alpha$, and the degree of freedom $\nu$. 
First, we estimate the cutoffs, i.e., the censoring thresholds, by taking the seasonality into account. Similar to \citet{sun2015stochastic}, the estimated cutoff $\hat{u}_t(s)$ is chosen to be the  quantile $q_t(s)$, corresponding to the probabilities $1-{O}_t(s)$, where $O_t(s)$ is the precipitation occurrence that takes a zero or one value and is fitted by logistic regression at each site $s$ with the binary time series data. The estimated occurrence is denoted by $\hat{O}_t(s)$. Then, the cutoff is estimated as $\hat{q}_t(s)$, the marginal sample quantile of $Y_t(s)$ corresponding to the probability $1-\hat{O}_t(s)$. Since $Y_t(s)$ is always larger than the real precision limit $u_r$, the cutoffs are not supposed to be smaller than $u_r$. Therefore, we have $\hat{u}_t(s)=\max\{\hat{q}_t(s),u_r\}$.  The covariates include harmonic terms for day-of-year seasonality. Here, we assume that
\begin{equation}
{
\begin{array}{r@{}l}
 \mathrm{logit}[&\mathbb{P}\{O_t(s)=1\}]=\sum_{j=1}^{H}\left\{\gamma_{1j}(s)+\gamma_{2j}(s)\sin\left(2\pi j \frac{d(t)}{365}\right) + \gamma_{3j}(s)\cos\left(2\pi j \frac{d(t)}{365}\right)\right\},
 \end{array}
\label{eq:ut}
}
\end{equation}
where $d(t)\in \{1,\dots, 365\}$ denotes the day within each year,  and the value of $H$ is chosen by Akaike information criterion (AIC \citep{akaike1998information}).%$h(t)\in \{1,\dots, 24\}$ denotes the hour within each day, 

%Second, we propose to use information inside the data to parameterize the heteroscedastic standard deviations. In specific, we assume $\sigma_t=b_0+b_1Y_{t-1}(s)$, where $b_0, b_1\geq0$ to avoid negative variance. 
%This parameterization is motivated by autoregressive conditionally heteroscedastic (ARCH) model, which is widely used in predicting high-frequency data such as wind power and stock-return \citep{engle1982autoregressive, bollerslev1986generalized, taylor2009wind}. In the ARCH(1) model, $\sigma^2_t(s)=b_0+b_1\varepsilon^2_{t-1}(s)$. But due the censoring mechanism, $\varepsilon^2_{t-1}(s)$ is hard to be observed and estimated. Therefore, we directly use the nonnegative data in the past, $Y_{t-1}(s)$, to model $\sigma_t$.  The heteroscedastic standard deviation have large influence on our model. $b_0$ mainly affects $Y_t(s)$ if $Y_{t-1}(s)$ is zero. Assuming that $\textbf{Y}_{t-1}=\textbf{0}$ and $b_0\rightarrow0$ as an extreme case, we have $\mathbb{P}(Y_t(s)=0|\textbf{Y}_{t-1}=\textbf{0})=F\left\{\frac{u_t(s)}{b_0}\right\}\rightarrow1$, indicating infinite dry spell length. When $b_0$ increases, there is more probable to have rain and the increase of the probability depends on other parameters. On the other hand, $b_1$ affects $Y_t(s)$ when $Y_{t-1}(s)$ is positive. Larger $b_1$ typically means longer wet spell length and higher intensity.

%In our application, the autoregression matrix is $8\times 8$ ($N=8$), which makes the estimation of the autoregressive parameters computationally difficult. 

To quantify the spatial dependence and make the estimation of the autoregressive parameters computationally feasible, we further parameterize the $N\times N$ matrix $B$ using the idea in \citet{sigrist2012dynamic}. Since we do not observe nonstationarity and anisotropy, we consider a simple parametrization of \citet{sigrist2012dynamic} that models $\beta_{ij}$ by a spatial covariance functions of Whittle-Mat{\'e}rn type i.e., $\beta_{ij}(\phi,\rho)=(\phi d_{ij}/\rho)K_{1}(d_{ij}/\rho)$, where $K(\cdot)$ is the modified Bessel function of the second kind, $\phi, \rho>0$ are scaling parameters, and $d_{ij}$ is the distance between locations $i$ and $j$. Thus we account for the spatial dependence and assume that faraway sites are less correlated. Furthermore, to make sure the process is temporally stationary, the largest eigenvalue of matrix $B$ should be less than 1. Here, we use a sufficient condition for $B$ by constraining $\phi/\rho<\frac{1}{N\max_{ij}(d_{ij})}$. Note that the condition only holds if $B$ is a valid covariance matrix.

The final unknown parameters in the \eqref{eq:lik} are $\boldsymbol{\theta}= (\phi, \rho, b_0, b_1,\alpha,\nu)^T$. Although we reduce the number of unknown parameters to six, the optimization of the likelihood that can only be achieved numerically is still potentially unstable. Thus, we consider several different numerical optimization methods: two derivative-free algorithms (COBYLA and Nelder-Mead) and two derivative-based algorithms (BFGS and CG). Many literatures point out that the \texttt{optim} function in R is not numerically stable, especially when a re-parameterization exists \citep{mullen2014continuous, nash2011unifying, nash2014best}. Therefore, we employ two recently developed R packages, \texttt{nloptr} \citep{johnson2014nlopt} and \texttt{Rcgmin} \citep{Nash:2014aa}, as a substitution of \texttt{optim}. To make sure that the optimization reaches the global maximum, we use different optimization algorithms with multiple sets of initial values until we get the same optimized values. The \texttt{sn} packages \citep{azzalini2011r} were used to evaluate $f_{\mathcal{SST}}(\cdot)$ and $F_{\mathcal{SST}}(\cdot)$. We also notice that the estimation of the degree of freedom $\nu$ is typically not numerically stable as in other similar problems, one can evaluate the likelihood over a sequence of values of $\nu$. The best $\nu$ can be selected according to the maximized likelihood function.%We choose $\nu=3$ to represent heavy tail, $\nu=7$ to represent moderate tail, and $\nu=20$ to represent light tail, because a value of $\nu$ smaller than 3 leads to an infinite mean or variance, and a very large $\nu$ is too close to the Gaussian case and significantly increases the computational burden. 

%%%%%%%%%%%%%%%%%%
%       Simulation Study
%%%%%%%%%%%%%%%%%%

\section{Simulation studies}\label{sec:sim}

We designed a simulation study to validate the inference procedure introduced in Section \ref{sec:inf} and compare to the original censored VAR models proposed by \citet{koch2015frailty} using the Gaussian dynamics. %However, a true comparison is hard to make. Few researchers have developed toolboxes or packages for replicating their generators' results, and most of the models are fundamentally different. Thus, a reasonable comparison can only be made with censored VAR models, where the independent random error $z_t(s)$ has different distribution. We consider the case when $z_t(s)$ is Gaussian distributed, as the SPG is a variant of \citet{koch2015frailty}. We show that when the true model is highly right-skewed and heavy-tailed, e.g., in the case of high-frequency rainfall, a Gaussian error is not sufficient to reproduce the true rainfall pattern, even with a heteroscedastic standard deviation. 
We generate multiple synthetic datasets from the censored VAR model \eqref{eq:pg} at the eight locations shown in Figure \ref{fig:rain_raw}, where $z_t(s)$ follows a skew-t distribution. By choosing different skewness parameter $\alpha$ and the degree of freedom $\nu$, the generated data can approximate the \citet{koch2015frailty} as a special case by letting $\alpha=0$ and $\nu$ be a large value. In the simulation, we set $\nu=20$ to be large and two other different degrees of freedom $\nu=3, 7$ for comparison. In addition, we set skewness parameter, $\alpha=5$ to represent the skewed data and set $(T, N, \phi, \rho, b_1, b_2)=(10000, 3, 1/3,1,0.5,0.5)$. The sample sizes, $T$ and $N$, are the same as the application in \citet{koch2015frailty} for comparison reason and other parameters are similar to the estimated values from the Lausanne precipitation data in order to mimic a real application. %As we mentioned when describing the inference procedure, the cutoffs are estimated separately. To prevent the cutoff values from influencing the estimation, we set the cutoffs to be the same as the estimated values from the real application as well.

%Figure S1 shows one realization at station CSS with six different model settings. It gives us a snapshot of the simulated data compared with the real data in Figure \ref{fig:rain_raw}.  We see that a smaller degree of freedom provides more extreme values, and larger skewness parameters result in denser heavy rains. %Despite having different lengths, the simulated time series data in Figure \ref{fig:alpha} represent the different scenarios of real rainfall in Figure \ref{fig:rain_15min} very well. For instance, in Figure \ref{fig:alpha}, top-left graph representing a light-tailed and symmetric case is similar to the 30-second rainfall data recorded at the location denoted by CSS in Figure \ref{fig:rain_15min}, and the bottom-right graph representing a heavy-tailed and highly skewed case looks similar to the rainfall data collected at the location denoted by unithequeO. 
We fit the models with skew-t and Gaussian dynamics to the synthetic datasets. The summary of estimated values based on 50 independent realizations are shown in Table S1 of the Supplements. With different dynamics, the optimized common factors, such as $b_0$ and $b_1$, are different as well. The results show that Gaussian dynamics tends to overestimate the $b_0$ and $b_1$ so that the overestimated variance will compensate the light tail of Gaussian distribution. 

Then, we generate 50 parametric bootstrap samples from the median of the fitted values and calculate the mean of the root-mean-squared errors (MRMSE) for the six scenarios. Specifically, the MRMSE is defined as $\frac{1}{50}\sum_{k=1}^{50}\sqrt{\frac{1}{NT}\sum_{t=1}^{T}\sum_{s=1}^N\left\{Y_t(s)-Y^B_{kt}(s)\right\}^2}$, where $Y_t(s)$ and $Y^B_{kt}(s)$ denote the synthetic data and the $k$-th bootstrap sample at time $t$ and location $s$, respectively. The results are shown in Table \ref{tab:sim}. 
  \begin{table}[h!]
    \centering
    \caption{\label{tab:sim} Summary of the simulation study results. The MRMSEs are presented as percentages (\%). The first row lists the parameter setups for $(\nu,\alpha)$. The second and third rows list MRMSE of the censored VAR generators with the skew-t and Gaussian dynamics, respectively. The last row shows the MRMSE ratio of the Gaussian to skew-t models.}
%\fbox{%
\begin{tabular}{c c c c c c c}
\toprule
   Scenarios $(\nu,\alpha)$ & $(3,0)$  &     $(3,5)$  &     $(7,0)$    &   $(7,5)$   &    $(20,0)$  &     $(20,5)$ \\
   \midrule
   Skew-t & $32.91$& $54.85$& $26.01$& $49.27$& $24.23$& $38.55$ \\
    Gaussian & $35.41$& $59.32$ &$27.03$& $52.30$ &$24.58$ &$39.70$  \\
    Ratio &$1.076$& $1.082$& $1.039$ &$1.061$& $1.014$& $1.030$\\
 \toprule
\end{tabular}
\end{table}
From Table 1, we see that when $\nu$ is large and $\alpha=0$, two models produce similar results. Even in this case, the MRMSE values of our model are slightly smaller than those of the model with the Gaussian dynamics since $\nu$ is still smaller than infinity.  In contrast, when $\alpha$ is positive and $\nu$ is small, the difference in MRMSEs is more significant, and the Gaussian model becomes less reliable. 
\begin{figure}[h]
\centering
         \includegraphics[width=\linewidth]{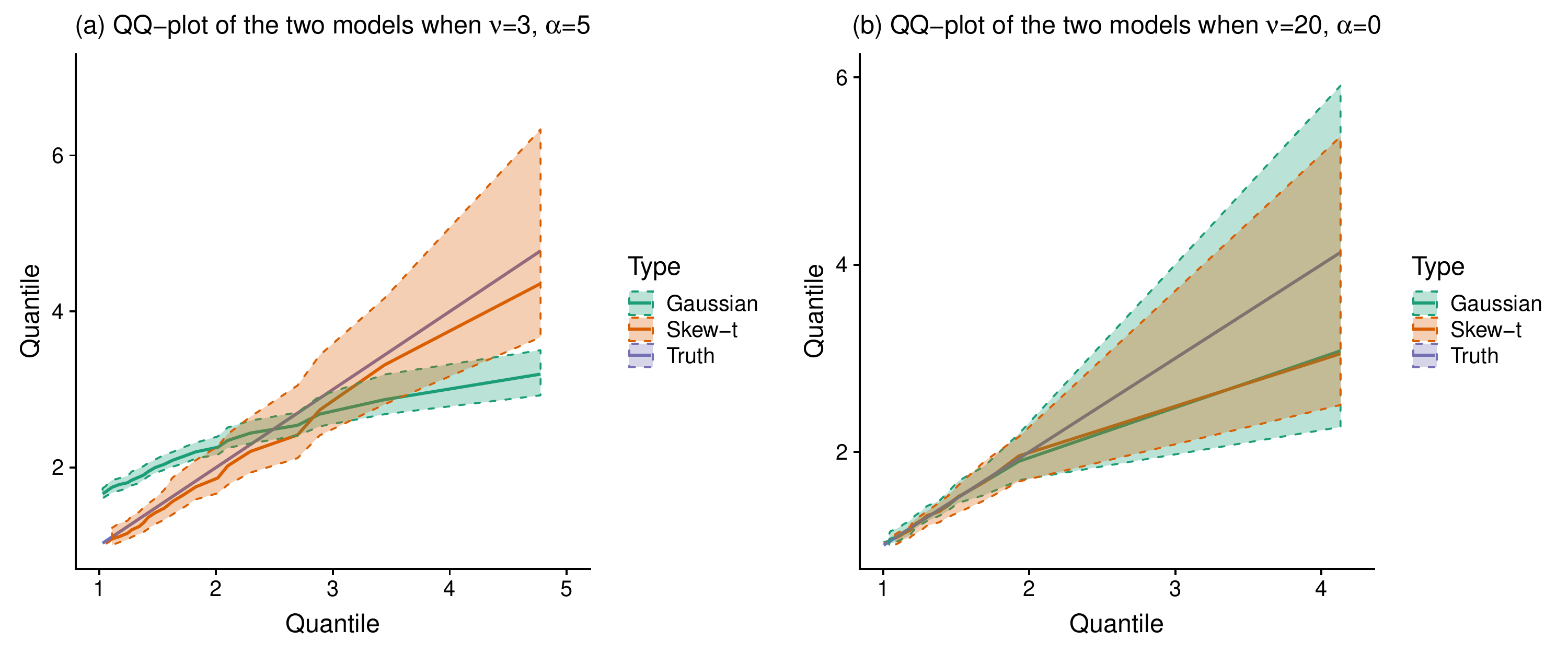}
         \caption{Quantile-quantile (QQ) plot between the synthetic (purple curve) and 50 parametric bootstrap samples using (a) a skew-t error (orange region) and (b) a Gaussian error (green region). The solid lines are the median curves and the shaded areas are the 95\% confidence intervals.}
         \label{fig:qqplot}
\end{figure}
To visualize the difference statistically, we draw quantile-quantile (QQ) plots between the synthetic data and the parametric bootstrap samples for two scenarios, $(\nu,\alpha)=(20,0)$ and $(\nu,\alpha)=(3,5)$. The results are shown in Figure \ref{fig:qqplot}. Not surprisingly, the Gaussian model cannot reproduce the heavy-tailed and right-skewed behavior that results from a large skewness parameter and a small degree of freedom. In both cases, our model with the skew-t errors successfully reproduces these statistical properties. 

Incorrectly assuming a Gaussian model for the data also affects the estimation of other important statistical properties of rainfall. As we mentioned, Gaussian model typically overestimates $b_0$ and $b_1$ of heavy-tailed data in order to produce a high rainfall intensity. However, as we explained in Section \ref{sec:inter}, a large $b_0$ and $b_1$ also leads to a low dry probability. Therefore, failing to correctly specify the degree of freedom will underestimate the dry probability as well, as shown in Figure S2 of the Supplements, where the distribution of the dry probability is obtained from the bootstrap samples. 
Therefore, only relying on the heteroscedastic standard deviations, without considering a right-skewed and heavy-tailed error term, is not enough to capture the statistical properties of high-frequency rainfall patterns and the stochastic simulations will not be realistic.

%%%%%%%%%%%%%%%%%%
%       Application
%%%%%%%%%%%%%%%%%%

\section{Application to Lausanne precipitation data}\label{sec:app}

The motivating data, as shown in Figure \ref{fig:rain_raw}, were collected by GAIA Lab, Institute of Earth Surface Dynamics (IDYST), the University of Lausanne in 2016 using eight Pluvimate acoustic rain gauges \citep{collister2008controls}. The detailed description of the Pluvimate rain gauges can be found in \citet{benoit2018stochastic}. This new instrument can measure precipitations at a $0.01$mm and 30 second resolution with the standard rain rate unit, mm/hr. From April 4th to April 14th, 2016, a total number of 230,400 observations were recorded by the network of eight rain gauges located within a radius of 1 km, as shown in Figure S2 in the Supplements. %by counting individual raindrops. Specifically, 1 count per epoch is equal to $0.01$ mm rain per $30$ seconds, and, thus, is equivalent to 1.2 mm/hr in terms of a common rain rate measure.  

Table \ref{tab:freq2} and Figure S3 in the Supplements show the distribution of the collected rainfall data. Most of the observations (92.4\%) were zeros, corresponding to no rain or dry events. Most of the rain intensities are below 5 mm/hr, while at some periods there are heavy rainfall with the rain rate around $10$ mm/hr. We also observe that the highest rain rate reached more than $30$ mm/hr. The QQ-plot also shows that these rainfall data are zero-inflated, right-skewed, and heavy-tailed and completely deviate from Gaussian distribution.

\begin{table}[h!]
\centering
\caption{Distribution of the collected rainfall data.}
\begin{tabular}{c|c c c c c c c c}
\toprule
   Rain rate (mm/hr)& $=0$  &    $(0,1.2]$  &    $(1.2, 5]$    &  $(5,10]$   &  $(10,33.6]$\\
 \midrule
              Percentage              & $92.40\%$& $5.55\%$&$1.84\%$ & $.19\%$&$.02\%$\\
\bottomrule
\end{tabular}
\label{tab:freq2}
\end{table}

We fitted the time-varying cutoffs as described in Section \ref{sec:inf} with the lowest positive rate rate $u_{r}=1.2$ using the \texttt{glm} package in R and fitted the other parameters by the methods mentioned in the Section \ref{sec:inf}. The fitted results are shown in Table \ref{tab:app}, along with the results of the Gaussian model for comparison purposes. We can see that the maximum likelihood estimators (MLEs) of both the degree of freedom $\nu$ and the skewness parameter $\alpha$ are small. Similar to the simulation study, the values of $b_0$ and $b_1$ estimated from the Gaussian model are larger than those estimated from the skew-t model.%When no rain occurs, e.g., the periods April 7 to April 8 and April 10 to April 12, the corresponding threshold $\hat{u}_t(s)$ is relatively large. 

  \begin{table}
    \caption{\label{tab:app}Maximum likelihood estimators (MLEs) from the censored VAR model with skew-t and Gaussian errors.}
  \centering
\begin{tabular}{c|c| c| c| c}
\toprule
   Parameters &\multicolumn{2}{c|}{Skew-t Errors} & \multicolumn{2}{c}{Gaussian Errors} \\
   &Est& 95\%CI & Est & 95\%CI\\
   \midrule
   $\phi$ & $.146$ &  $(.143,.149)$ &  $.163$ &  $(.158,.168)$   \\
   $\rho$ &  $1.395$ &  $(1.211,1.610)$ &  $0.879$ &   $(0.786,0.984)$    \\
   $b_0$ & $.457$ &  $(.451,.463)$ &  $.474$ &   $(.470,.478)$    \\
   $b_1$ & $.257$ &  $(.250,.271)$ &  $.281$ &   $(.273,.290)$    \\
   $\alpha$ & $.034$ &  $(.022,.053)$ &   &       \\
   $\nu$ &4&&&\\
   \toprule
\end{tabular}
\end{table}
\subsection{Conditional rainfall generation}
\begin{figure}[h!]
\centering
         \includegraphics[width=\linewidth]{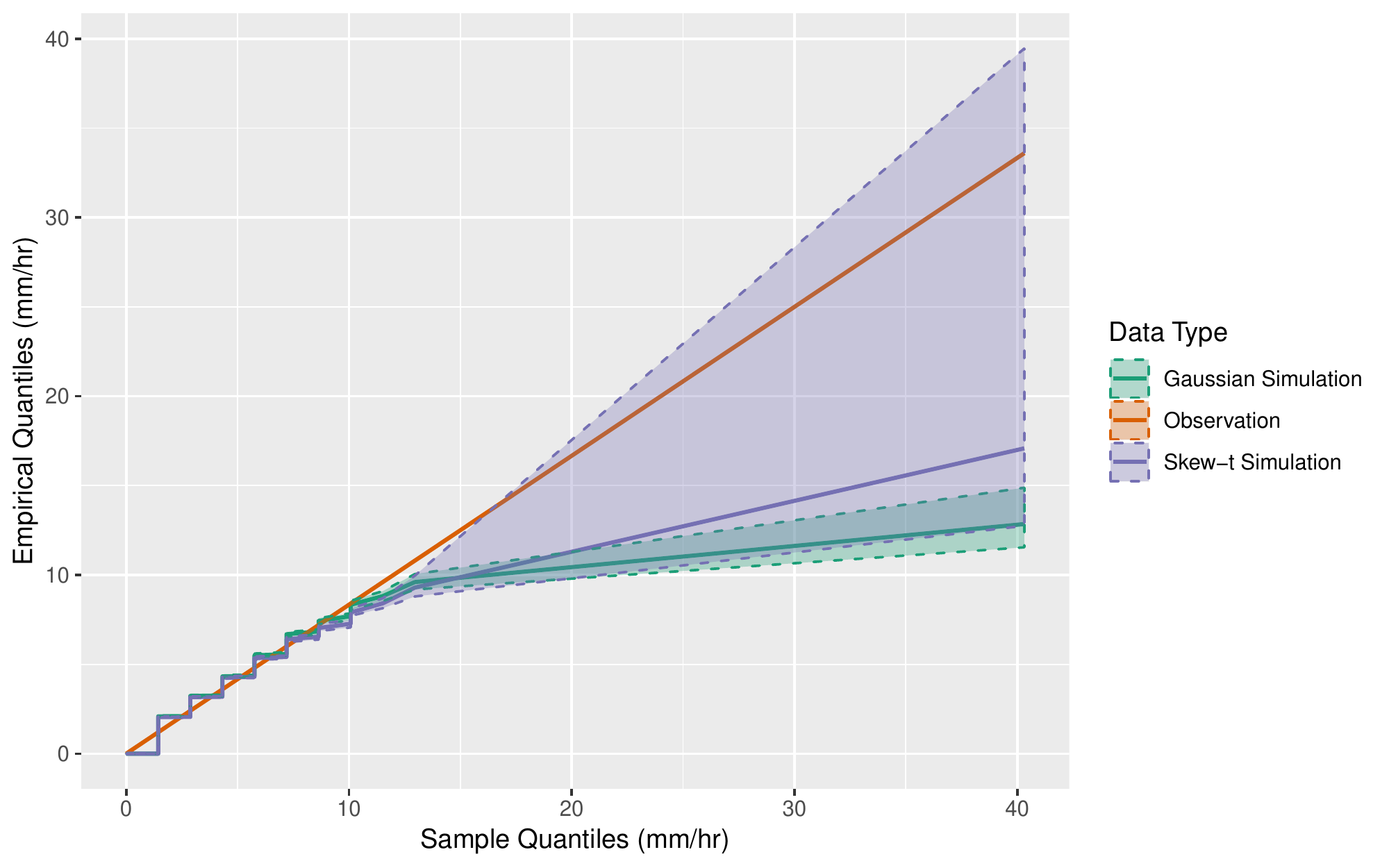}
         \caption{Quantile-quantile (QQ) plot between the observed rainfall (purple curve) and 50 parametric bootstrap samples using a skew-t error (orange region) and a Gaussian error (green region). The solid lines are the median curves and the shaded areas are the  95\% confidence intervals.}
         \label{fig:qqapp}
\end{figure}

From the fitted model, we can generate 50 parametric bootstrap samples of the high-frequency rainfall. Each sample can be viewed as a synthetic realization of the 30-second rainfall data. To examine the similarity of the simulated samples to the real data, we use a one-step-ahead conditional simulation, which generates data conditional on observations one step in the past. The main goal here is to reproduce the statistical properties of the real data, we use a QQ plot shown in Figure \ref{fig:qqapp}, to compare the real data with the bootstrap samples from the two models. From the results, both models can reproduce the low-intensity rainfall well statistically. However, in terms of the heavy rainfall (defined as having an intensity larger than 10 mm/hr), the 95\% confidence band of the Gaussian model significantly underestimates the precipitation, whereas the skew-t model captures the heavy tail very well. %However, the uncertainty of the heavy rainfall generated by the skew-t model is very large, reflecting the high variability of the high-frequency rainfall.

\begin{figure}[h]
\centering
         \includegraphics[width=\linewidth]{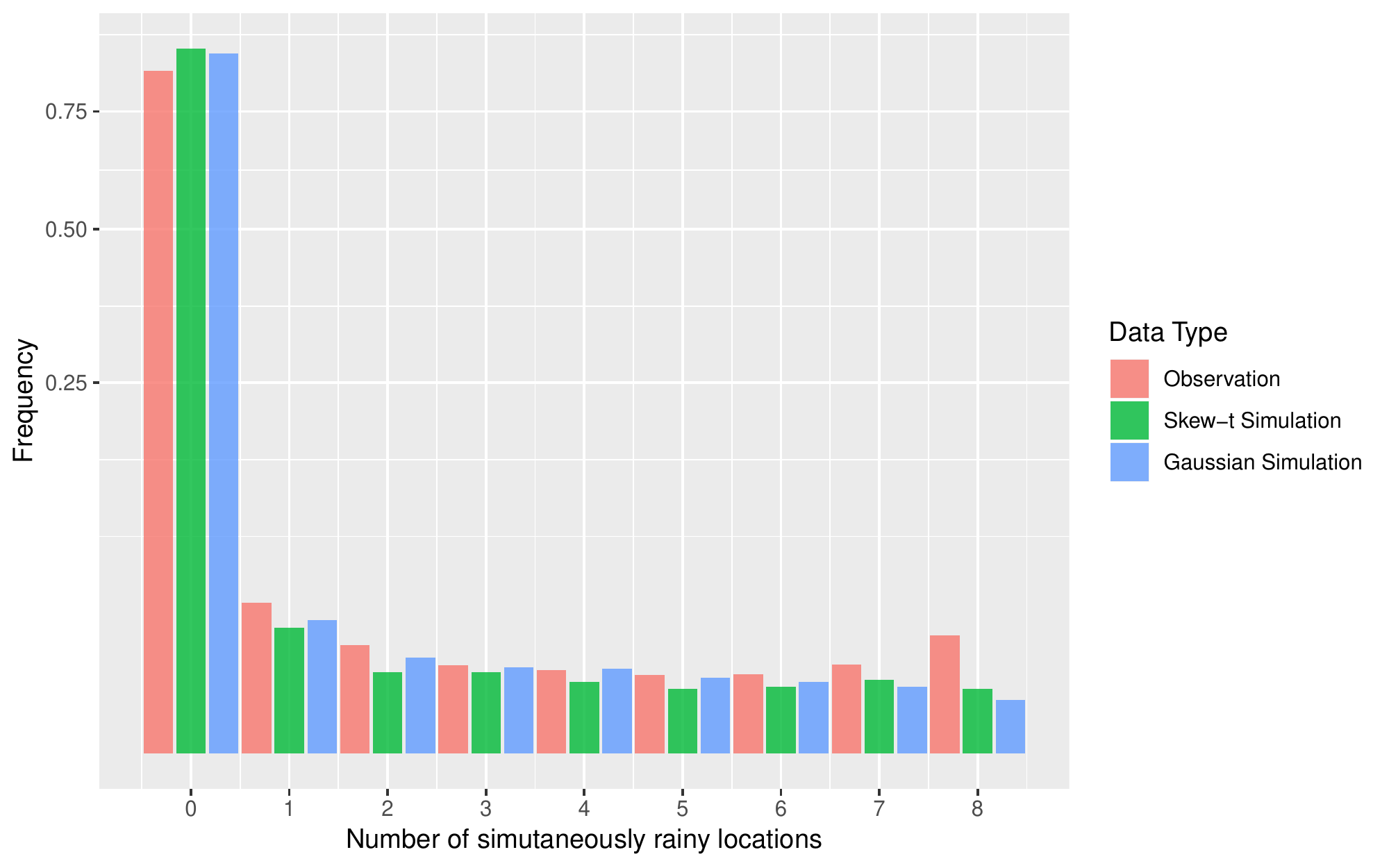}
         \caption{Histogram showing the simultaneously rainy locations. Each bar represents how many corresponding events happen in time. The y-axis is shown in the square root scale.}
         \label{fig:spatial}
\end{figure}

The spatio-temporal patterns of rainfall occurrence are assessed by the rain concurrences and dry probability. Figure \ref{fig:spatial} shows the histogram of simultaneously rainy locations for both the observed and simulated rainfall data, from which we can see the spatial dependences among multisite rainfall occurrences. The simulated histograms combines all of the 50 parametric bootstrap samples. The eight locations are usually all dry. In this case, the simulated data slightly overestimate the counts. In contrast, when at least one location is rainy, the simulated data have fewer counts than real observations, which means the spatial correlation is slightly underestimated. Overall, the rain concurrences are reproduced reasonably well by our model.
The temporal dependence of rainfall occurrence can be reflected by the dry/wet spell length and conditional dry/wet probability. The dry/wet spell length is less important for the high-frequency data of only ten days and thus we only show the conditional probabilities.
Table \ref{tab:condiprob} displays the four different conditional probabilities. Overall, the simulated data can reproduce the conditional dry/wet probability well. However, the simulated data tend to underestimate the probability of transition between wet and dry. This is an expected result of the hard thresholding effect. Nevertheless, it does not affect much on the overall distribution of the rainfall intensity. If it is crucial to accurately reproduce the dry or wet probability, a separate model for rainfall occurrences might be a better choice. 
%Figure \ref{fig:length} shows the stacked histogram of dry spell lengths from the simulated data and the true observations. The simulated data can capture the long and short dry periods well. In general, however, the model tends to underestimate the medium dry spell length. 

  \begin{table}
    \caption{\label{tab:condiprob}Conditional probabilities from the observed rainfall data and the mean ($\pm1.96\times$standard deviation) of the conditional probabilities from 50 parametric bootstrap samples generated using a skew-t and Gaussian model.}
  \centering
%\fbox{%
\begin{tabular}{c|c c c c}
\toprule
    &Wet$|$Wet & Dry$|$Wet & Wet$|$Dry & Dry$|$Dry\\
   \hline
   Skew-t Dynamics & .976($\pm.0045$)& .024($\pm.0045$) &.024($\pm.0045$)& .976($\pm.0045$)  \\
   Gaussian Dynamics & .974($\pm.0062$)& .026($\pm.0062$)&.026($\pm.0062$)& .974($\pm.0062$)  \\
   Observation &.971& .029& .029& .971    \\
   \bottomrule
\end{tabular}
\end{table}

%\begin{figure}[h]
%\centering
 %        \includegraphics[width=.49\linewidth]{length.pdf}
  %       \includegraphics[width=.49\linewidth]{wetlength.pdf}
   %      \caption{Quantile-quantile (QQ) plots of the dry and wet spell lengths for the observed rainfall data (orange curve) and 50 bootstrap samples using a skew-t error (green region). The solid lines are the median curves and the shaded areas are the 95\% confidence intervals.}
%         \label{fig:length}
%\end{figure}

Compared to the Gaussian model, the major advantage of the skew-t model is that it can reproduce both dry events and heavy rainfall. However, since the Gaussian and skew-t models we considered use the same spatio-temporal modeling strategy, i.e., the VAR model with a time-varying threshold, the spatio-temporal properties of the rainfall occurrences do not show much difference. 

\subsection{Short-term rainfall prediction}
A SWG is not meant for prediction but generate multiple realizations with similar statistical properties of the real data. However, we can provide some predictive information by the mean of quantiles of multiple realizations. We give an example here to illustrate the performance of our model in prediction, compared to a multivariate LSTM model as a popular deep learning-based model for time series prediction. We use the last day of rainfall ($5,760$ points) at each location as the testing set and other data as the training set. We re-fit our model using the training data and fit the LTSM model using one hidden layer of $100$ LSTM units with rectified linear unit (ReLU) activation, Adam optimization algorithm, and mean squared error (MSE) as the loss function. The LSTM model is run on Keras functional API with 100 epoch in Python.

%Figure \ref{fig:pred_compare} shows the predicted values on the entire testing set. 

\begin{figure}[h]
\centering
         \includegraphics[width=.9\linewidth]{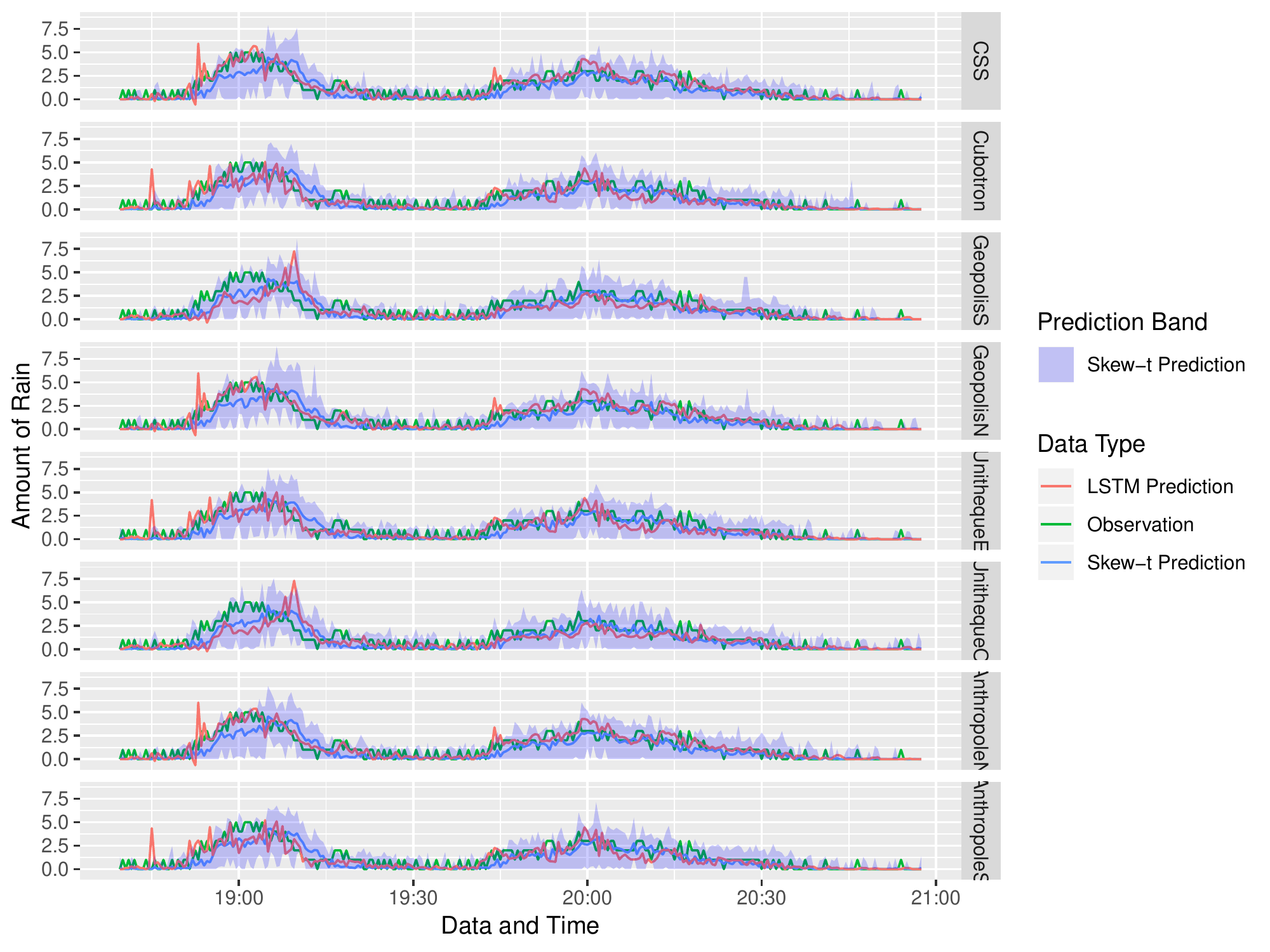}
         \caption{The performance of predicted rain rate (mm/hr) (with uncertainties) during a heavy rainfall period on April 13, 2016. The blue line and blue band are the mean prediction and $95\%$ prediction band from our model, respectively. The red line is the prediction from LSTM model. The green line is the last day of rainfall observation as the truth.}
         \label{fig:pred_compare}
\end{figure}

As a precipitation generator, our model can simultaneously generate 50 realizations and we use the mean of 50 realizations as our predictor in the comparison. We also compute the $95\%$ prediction band by the quantiles of the 50 realizations to show the uncertainties. In terms of the MSE, the result of our model (MSE$=0.099$) outperforms the LSTM (MSE$=0.126$). However, if we pick a single realization of our model, the MSE of LSTM is better than any of the realizations  (MSE$=0.204$ based on the best MSE of single realization). To visualize the results, in Figure \ref{fig:pred_compare}, we show one heavy rain event in the evening of April 13 and $95\%$ prediction band from our model as there is almost no rain at other time. The results show that the 95\% prediction band covers the observed rainfalls. Also, the prediction of LSTM is more variable with a few obvious peaks far from the observed values. Instead, our predictor is more stable since it collects the mean of 50 realizations.%%

From the results, two main differences between LSTM predictor and our stochastic precipitation generator are noticed, even though both of the models provide good prediction. Firstly, a SWG can serve as a predictor by generating multiple independent predictors, which can be used to quantify the uncertainties. But with regards to a single predictor, LSTM can perform better. Secondly, as many other deep learning methods, the prediction of LSTM is done via optimizing certain loss function, e.g. MSE, thus the results are hard to interpret than our model-based prediction. For example, the predicted values from LSTM model can be negative values. Although we can treat the negative value as zero by the ReLU activation function in the output layer, we lose some meaningful interpretations of the rainfall properties.

\section{Discussion}\label{sec:dis}
In this study, we proposed a new SWG for high frequency rainfall that is capable of reproducing large quantities of zeros and extreme values. This model utilizes skew-t random variables with a censoring mechanism to simultaneously drive both the occurrence and intensity of rainfall. By applying a VAR model for spatio-temporal dependence, the inference can be achieved simply by maximizing the likelihood function.

We applied the SWG to a rarely assessed and fine-scale precipitation dataset collected by an acoustic rain gauges every 30 seconds and spatially around $10-100$ meters. The results show that the SPG can generate high-intensity rainfall better than the baseline model with Gaussian errors. We also show that our SWG can be used for prediction purposes and its performance is comparable to LSTM, a modern deep learning method. Compared to the LSTM, our SWG can provide uncertainties and is more interpretable.

Although we only consider the VAR framework with a linear spatio-temporal dependence, the model can be generalized by considering other spatio-temporal dependences. One situation was discussed in \citet{tadayon2015bayesian}, where a Bayesian hierarchical model was used with only skew-normal error and purely spatial dependence. Another possibility is to use a censored multivariate skew-symmetric distribution. However, the associated inference might be challenging, since likelihood functions do not have a closed form as in our model. The main reason is that skew-symmetric random variables do not retain all of the desired properties when they become Gaussian or student-t random variables, e.g., they are not closed under convolution or conditioning. Thus, some existing results under certain assumptions, such as additive models, or solutions derived from conditional distributions, such as the Kriging predictor, would no longer hold in skew-symmetric cases.

We parameterize the autoregression matrix, which reduces the number of parameters while taking spatial information into account. Although such a parameterization fits our purpose well, some limitations exist. First, we selected a covariance function to construct the autoregression matrix, which implies that the single-site temporal correlation is homogenous across many locations. Second, the correlation matrix only allows for a positive correlation among locations, whereas an autoregression matrix could also have negative correlations. Last, we use a sufficient condition for the parameters in the spatial covariance function to satisfy the temporally stationary condition. However, it is more favorable to have a necessary and sufficient condition to ensure the stationarity. Thus if only less sites are assessed it is better to estimate the autoregression matrix in element-wise. 

%\newpage
%\section{Appendix}
\section{Acknowledgement}
This research was supported by King Abdullah University of Science and Technology (KAUST), Office of Sponsored Research (OSR) under Award No: OSR-2019-CRG7-3800. The dataset used in this study is provided by GAIA Lab, Institute of Earth Surface Dynamics (IDYST), the University of Lausanne. We acknowledge their efforts for collecting the data.

%\begin{supplement}
%\sname{Supplement A}\label{suppA}
%\stitle{Additional proofs, tables, and figures}
%\slink[url]{link}
%\sdescription{The supplement consists of the proof of some important equations and additional figures and tables.}
%\end{supplement}

\bigskip
\begin{center}
{\large\bf SUPPLEMENTARY MATERIAL}
\end{center}

\begin{description}

\item[Source code:] The source code in this work is available on the following repository: \\
https://github.com/aleksada/MultisiteHighFreqPG. (website)

\item[Proofs, figures, and tables:] The supplementary material consists of the proofs of Equations (4) and (5) and additional figures and tables.  (.pdf file)

\end{description}

\bibliographystyle{apalike}

\bibliography{reference}
\end{document}

\def\spacingset#1{\renewcommand{\baselinestretch}%
{#1}\small\normalsize} \spacingset{1}

%%%%%%%%%%%%%%%%%%%%%%%%%%%%%%%%%%%%%%%%%%%%%%%%%%%%%%%%%%%%%%%%%%%%%%%%%%%%%%

\if0\blind
{
  \title{\bf A Multi-Site Stochastic Weather Generator for High-Frequency Precipitation Using Censored Skew-Symmetric Distribution}
  \author{Yuxiao Li and Ying Sun}
    %and \\
    %Ying Sun \\
    %Statistics Program, King Abdullah University of Science and Technology}
  \maketitle
} \fi

\if1\blind
{
  \bigskip
  \bigskip
  \bigskip
\begin{center} {\Large{\bf Supplementary Material for ``Multi-Site High-Frequency Stochastic Precipitation Generator Using Censored Skew-Symmetric Distributions''}}
\end{center}
  \medskip
} \fi

\spacingset{2} % DON'T change the spacing!
\setcounter{figure}{0}
\renewcommand{\thefigure}{S\arabic{figure}}
\setcounter{table}{0}
\renewcommand{\thetable}{S\arabic{table}}

\section{Proofs}
\subsection{The equivalence between Equation (2) and Equations (3) and (4)} 
Equation (3) can be derived directly from Equation (2) by letting ${X}_{t}(s)=\boldsymbol{\beta}_s'\textbf{Y}_{t-1}(s)+\varepsilon_t(s)$. Further, denote Equation (3) by $\textbf{Y}_t=g(\textbf{X}_{t})$, then ${X}_{t}(s)=\boldsymbol{\beta}_s'\textbf{Y}_{t-1}+{\varepsilon_t}(s)=\boldsymbol{\beta}_s' g(\textbf{X}_{t-1})+{\varepsilon_t}(s)$. We note that each element $\varepsilon_t(s)$ follows skew-t distribution with mean zero and variance $\sigma_t$, which is linearly associated with a standard skew-t random variable for any $t$ and $s$. Therefore, using the conditioning representation of the standard skew-t random variable \citep{azzalini2013skew}, we have 
$\delta_t(s)=\begin{cases}
\begin{array}{c}
V_t(s),\\
-V_t(s),
\end{array} & \begin{array}{c}
W_t(s)>0 ,\\
W_t(s)\leq 0,
\end{array}\end{cases}$ where $V_t(s)$ and $W_t(s)$ are correlated standard t random variables. Since $\varepsilon_t(s)$ is a linear transformation of $\delta_t(s)$, we have ${\varepsilon_t}(s) ={c}_t(s)+ \begin{cases}
\begin{array}{c}
{Z}_t(s),\\
-{Z}_t(s),
\end{array} & \begin{array}{c}
{W}_t(s)>0 ,\\
{W}_t(s) \leq 0,
\end{array}\end{cases}$ with a fixed value ${c}_t(s)$.  Therefore, we can obtain (4) by combining terms. \qed

\subsection{The likelihood function in Equation (5)} 

The proof is an extension of the related methods in \citet{koch2015frailty}, as we generalize the error term to a non-Gaussian distribution. 

Denote the density function of $\{Y_t(s);t=1,\dots,T, s=1,\dots,N\}$ by $h(Y_t(s);t=1,\dots,T, s=1,\dots,N)$. Due to the Markov property of the VAR(1) process, the density can be written as 
$h(Y_t(s);t=1,\dots,T, s=1,\dots,N)=\prod\limits_{t=2}^{T}h(\textbf{Y}_{t}|\textbf{y}_{t-1})$. Since $z_t(s)$ are mutually independent for any $t$ and $s$, $u_t(s)$ is given, and $\sigma_t(s)$ only depends on $\textbf{Y}_{t-1}$, we can conclude that, conditional on $\textbf{Y}_{t-1}$, $Y_t(s)=(\boldsymbol{\beta}_{s}'\textbf{Y}_{t-1}+{\varepsilon_t}(s))\mathbbm{1}_{\{\boldsymbol{\beta}_{s}'\textbf{Y}_{t-1}+{\varepsilon_t}(s)>{u}_t(s)\}}$ as a function of $z_t(s)$ are mutually independent for any $t$ and $s$. Therefore, we have $h(Y_t(s);t=1,\dots,T, s=1,\dots,N)=\prod\limits_{t=2}^{T}\prod\limits_{s=1}^{N}h({Y}_{t}(s)|\textbf{y}_{t-1})$. 

Due to the censoring mechanism, the log-density of $Y_t(s)$, $\ell$, is a mixture of discrete part $\ell_d$ and continuous part $\ell_c$. When $y_t(s)>0$, $Y_t(s)=\boldsymbol{\beta}_{s}'\textbf{y}_{t-1}+(b_0+b_1\bar{Y}_{t-1})Z_t(s)$, and $Z_t(s)$ follows a skew-t distribution with density $f_{\mathcal{SST}}(x)$. By changing a variable, in this case, we have
\[\log\{h({Y}_{t}(s)|\textbf{y}_{t-1})\}=\left[\log\left\{f_{\mathcal{SST}}\left(\frac{y_t(s)-\boldsymbol{\beta}_{s}'\textbf{Y}_{t-1}}{b_0+b_1\bar{Y}_{t-1}}\right) \right\}-\log\left\{b_0+b_1\bar{Y}_{t-1}\right\}\right].\]
When $y_t(s)=0$, $\mathbb{P}\{y_t(s)=0|\textbf{y}_{t-1}\}=\mathbb{P}\{y_t(s)\leq u_t(s)|\textbf{y}_{t-1}\}$. By writing $Y_t(s)$ as a function of $Z_t(s)$, we can get that
\[\log \mathbb{P}\{y_t(s)=0|\textbf{y}_{t-1}\} = \log\left\{F_{\mathcal{SST}} \left(\frac{\hat{u}_t(s)-\boldsymbol{\beta}_{s}'\textbf{y}_{t-1}}{b_0+b_1\bar{Y}_{t-1}}\right) \right\}\]

Therefore, we have
\begin{align*}
&\log\{h({Y}_{t}(s)|\textbf{y}_{t-1}\}\\
&=\sum\limits_{t=2}^{T}\sum\limits_{s=1}^{N}\ell_c\mathbbm{1}_{\{y_t(s)>0\}}+\ell_d\mathbbm{1}_{\{y_t(s)=0\}} \\
&=\sum_{t=2}^T\sum_{s=1}^N  \mathbbm{1}_{\{y_t(s)>0\}}\left[\log\left\{f_{\mathcal{SST}}\left(\frac{y_t(s)-\boldsymbol{\beta}_{s}'\textbf{y}_{t-1}}{b_0+b_1\bar{Y}_{t-1}}\right) \right\}-\log\left\{b_0+b_1\bar{Y}_{t-1}\right\}\right] \\
&\hspace{5mm} +\sum_{t=2}^T\sum_{s=1}^N\mathbbm{1}_{\{y_t(s)=0\}} \log\left\{F_{\mathcal{SST}} \left(\frac{\hat{u}_t(s)-\boldsymbol{\beta}_{s}'\textbf{y}_{t-1}}{b_0+b_1\bar{Y}_{t-1}}\right) \right\}.
\end{align*}
\qed

\bibliographystyle{apalike}
\bibliography{reference}
\section{Additional Tables and Figures}
%\subsection{Tables}[h!]

%\subsection{Figures}

%\begin{figure}[h!]
%\centering
 %        \includegraphics[width=0.7\linewidth]{vis_nu_alpha.pdf}
 %        \caption{Simulated rainfall at CSS with different skewness parameters and degrees of freedom}
  %       \label{fig:alpha}
%\end{figure}

\begin{figure}[h!]
\centering
         \includegraphics[width=0.65\linewidth]{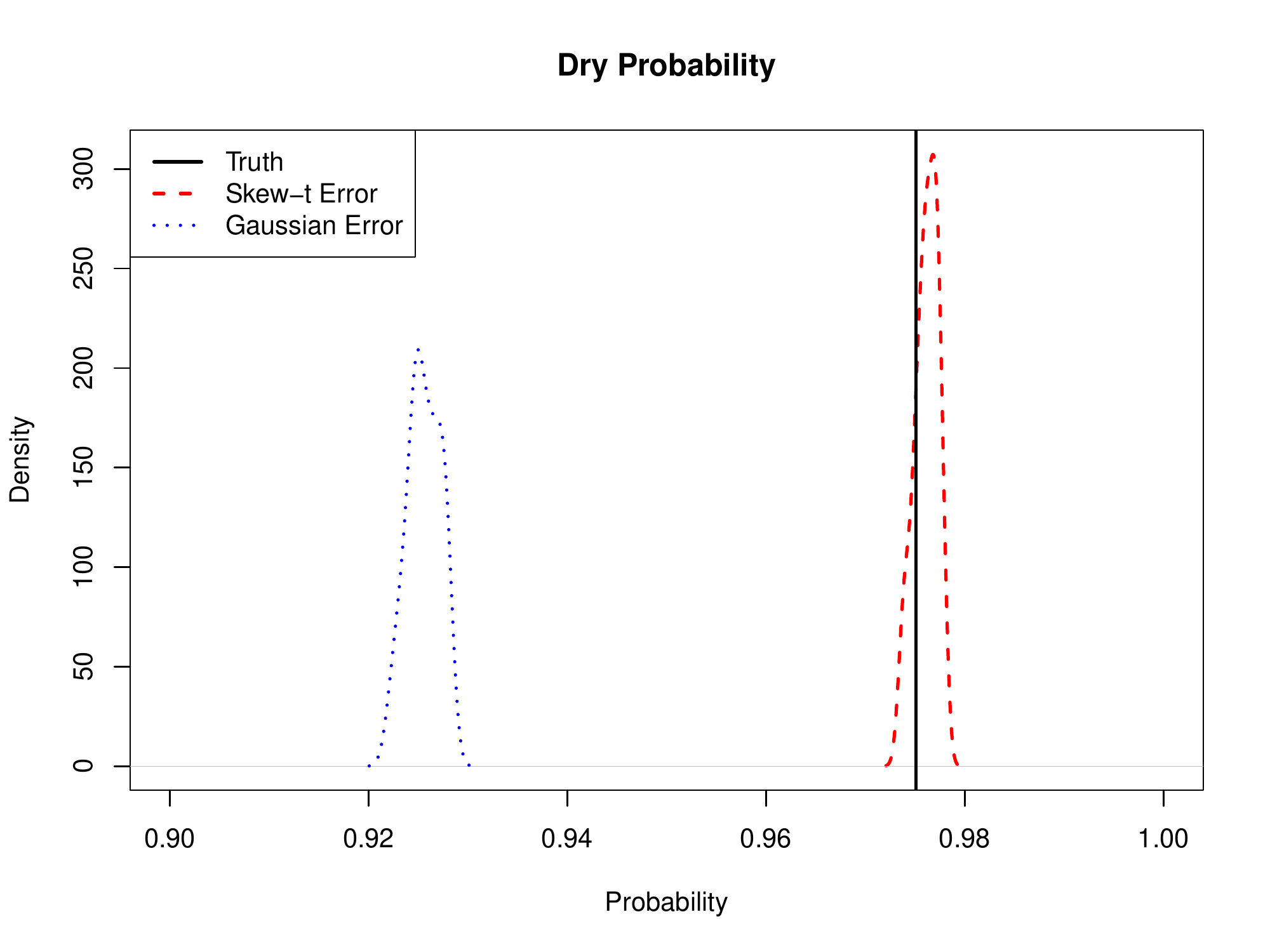}
         \vspace{-2mm}
         \caption{The dry probabilities from the synthetic dataset and 50 parametric bootstrap samples using skew-t and Gaussian models. The black line represents the truth from the synthetic data; the red dashed and blue dotted curves are the empirical density of the dry probability from 50 parametric bootstrap samples using Gaussian and skew-t models, respectively.}
         \label{fig:dry}
\end{figure}

\begin{figure}[h!]
\centering
         \includegraphics[width=.6\linewidth]{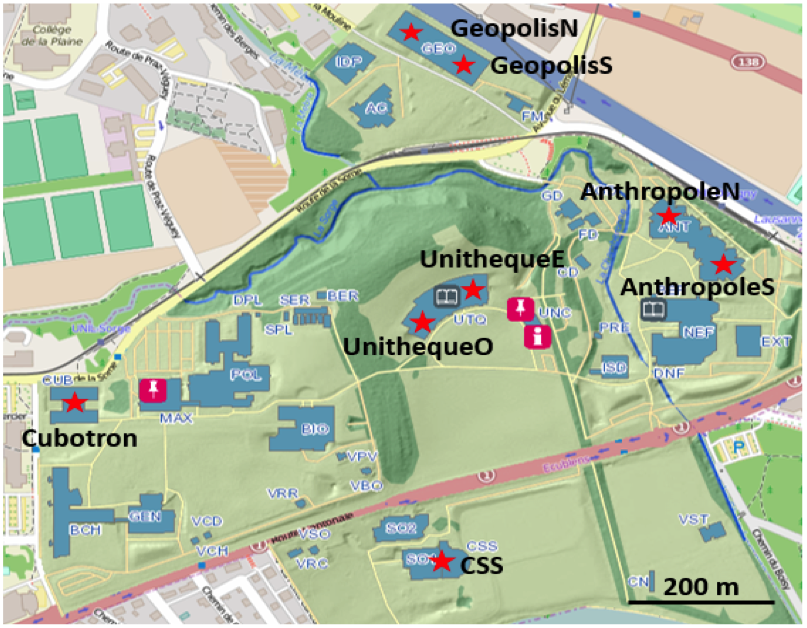}
         \caption{Map of the University of Lausanne showing precipitation measurement locations (red stars).}
         \label{fig:location}
\end{figure}

\begin{figure}[h!]
\centering
                  \includegraphics[width=.7\linewidth]{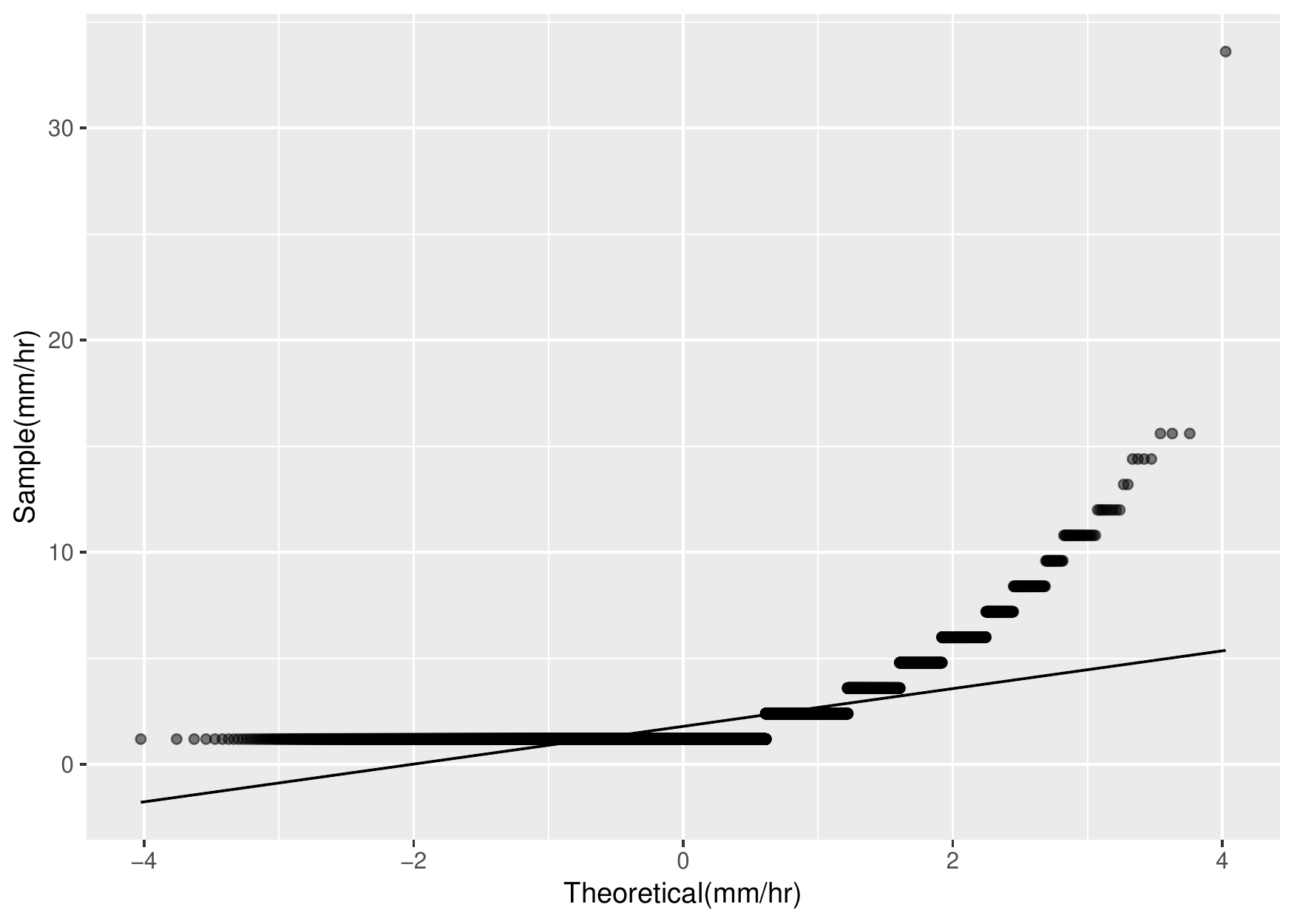}
         \caption{Distribution of rainfall data visualized using a normal quantile-quantile (QQ) plot (only positive values). The figure implies a right skewness of the rain distribution.}
         \label{fig:eda}
\end{figure}

\begin{table}[h!]
\centering
\caption{Summary of the simulation study results. Every four rows list the estimated mean and standard deviation of one parameter from six scenarios for $(\nu,\alpha)$. Given a parameter, the first two rows show the results from the proposed skew-t model, and the last two rows provide the results from Gaussian model for our comparison.}
\begin{tabular}{c | c|c c c c c c}
\toprule
   Parameters &Scenarios $(\nu,\alpha)$ & $(3,0)$  &     $(3,5)$  &     $(7,0)$    &   $(7,5)$   &    $(20,0)$  &     $(20,5)$ \\
   \midrule
   $\phi (=.25)$ & Skew-t (mean) & $.260$& $.252$& $.229$& $.230$& $.234$& $.236$ \\
    &   Skew-t (sd) & $.031$& $.037$ &$.031$ &$.039$& $.048$ &$.035$   \\
       & Gaussian (mean) & $.119$& $.007$& $.160$& $.051$& $.211$& $.124$ \\
    &  Gaussian (sd)& $.064$& $.016$ &$.036$& $.040$ &$.046$ &$.036$  \\
    \midrule
   $b_0(=.5)$ & Skew-t (mean) & $.491$& $.509$& $.491$& $.507$& $.496$& $.507$ \\
    &   Skew-t (sd) & $.018$& $.020$ &$.010$& $.020$ &$.007$ &$.021$  \\
       & Gaussian (mean) & $.624$& $.825$& $.539$& $.671$& $.513$& $.620$ \\
    &  Gaussian (sd)& $.049$& $.109$ &$.009$& $.010$ &$.006$ &$.008$  \\
    \midrule
   $b_1(=.5)$ & Skew-t (mean) & $.485$& $.481$& $.530$& $.516$& $.483$& $.494$ \\
    &   Skew-t (sd) & $.097$& $.069$ &$.063$& $.051$ &$.059$ &$.054$  \\
        & Gaussian (mean) & $.387$& $.651$& $.534$& $.688$& $.480$& $.609$ \\
    &  Gaussian (sd)& $.207$& $.242$ &$.082$& $.099$ &$.056$ &$.070$  \\
    \midrule
   $\alpha$ & Skew-t (mean) & $.002$& $2.646$& $.059$& $4.020$& $.017$& $5.492$ \\
    &   Skew-t (sd) & $.128$& $44.487$ &$.188$& $30.631$ &$.273$ &$27.486$  \\
\bottomrule
\end{tabular}
\label{tab:sim2}
\end{table}

\vspace{1.5cm}
%\begin{figure}[h!]
%\centering
 %        ~~~~~~~~\includegraphics[width=.7\linewidth]{ut}
 %        \caption{Fitted series of the cutoff parameters $u_t(s)$ with seasonalities at eight locations. We consider both the the hourly and daily seasonalities.}
 %        \label{fig:ut}
%\end{figure}

% AOS,AOAS: If there are supplements please fill:
%\begin{supplement}[id=suppA]
%  \sname{Supplement A}
%  \stitle{Title}
%  \slink[doi]{10.1214/00-AOASXXXXSUPP}
%  \sdatatype{.pdf}" 
%  \sdescription{Some text}
%\end{supplement}